\documentclass[aps,prl,preprint,groupedaddress]{revtex4-1}
\usepackage{graphicx}
\usepackage{multirow}
\usepackage{longtable}
\newcommand{\upcite}[1]{\textsuperscript{\textsuperscript{\cite{#1}}}}
\begin{document}

\title{The ultra-low-frequency shear modes of 2-4 layer graphenes observed in their scroll structures at edges}

\author{Ping-Heng Tan}
\email{phtan@semi.ac.cn}
\author{Jiang-Bin Wu}
\author{Wen-Peng Han}
\author{Wei-Jie Zhao}
\author{Xin Zhang}
\affiliation{State Key Laboratory of Superlattices and Microstructures,
Institute of Semiconductors, Chinese Academy of Sciences, Beijing 100083,
China}
\author{Hui Wang}
\author{Yu-Fang Wang}
\affiliation{Department of Physics, Nankai University, Tianjin 300071,
China}

\date{\today}

\begin{abstract}
The in-plane shear modes between neighbor-layers of 2-4 layer graphenes (LGs) and the corresponding graphene scrolls rolled up by 2-4LGs were investigated by Raman scattering. In contrast to that just one shear mode was observed in 3-4LGs, all the shear modes of 3-4LGs were observed in 3-4 layer scrolls (LSs), whose frequencies agree well with the theoretical predication by both a force-constant model and a linear chain model. In comparison to the broad width (about 12cm$^{-1}$) for the G band in graphite, all the shear modes exhibit an intrinsic line width of about 1.0 cm$^{-1}$. The local electronic structures dependent on the local staking configurations enhance the intensity of the shear modes in corresponding 2-4LSs zones, which makes it possible to observe all the shear modes. It provides a direct evidence that how the band structures of FLGs can be sensitive to local staking configurations. This result can be extended to n layer graphene (n $>$ 4) for the understanding of the basic phonon properties of multi-layer graphenes. This observation of all-scale shear modes can be foreseen in other 2D materials with similar scroll structures.

\end{abstract}


\maketitle

\section{Introduction}
Single-layer Graphene (SLG) has high mobility and optical transparency, as well as flexibility, robustness and environmental stability\upcite{Geim-nm-07-graphene,bonaccorso-NP-2010-graphene}. Few-layer graphene (FLG) with less than ten layers each show a distinctive band structure\upcite{koshino-2009-SSC-electronic}. That has been making people be interesting in the physics of FLGs and their application in useful devices\upcite{zhu-2009-PRB-carrier,ye-2011-NAS-accessing,huang-2011-SMLL-appli}. There are different stacking ways for FLGs, such as AB, AA for bilayer (2LG), trilayer (3LG), four-layer graphene(4LG), or more, ABC for 3LG or more and other stacking configurations\upcite{lui-2010-NL-imaging}. The stacking configuration, conditioned by preparation method, correspond the band structure and other properties. One stacking configuration, discussed frequently recently, is twisted bilayer graphene\upcite{dos-PRL-2007-graphene}, which can be grown epitaxially on the SiC(000\={1}) surface\upcite{Berger-science-2006-grown,hass-2008-PRL-multilayer} or by Chemical Vapor Deposition (CVD) on Cu/Ni or other substrates\upcite{kim-Nature-2009-large,Reina-nl-2009-Large,ago-SMLL-2009-growth}. In twisted bilayer graphene, two Dirac electron gases are then coupled by a periodic interaction, with a large supercell, which can restore a Dirac-like linear dispersion with lower Fermi velocity than single layer graphene\upcite{dos-PRL-2007-graphene,ni-PRB-2008-reduction,trambly-NL-2010-localization}. It was found to have two Van Hove singularities in its density of states\upcite{li-2009-NPhysi-observation}, which can resonant with incident laser with corresponding energy during Raman scattering process, making intensity of G peak extremely strong\upcite{havener-NL-2012-angle,kim-prl-2012-raman}. Sometime, there would be a few wire-like scroll structures at the edges of FLGs sample, which prepared by Micro-mechanical exfoliation\upcite{meyer-Nature-2007-structure,Jiang-2009-NL}. Abnormal stacking configurations, similar to twisted bilayer graphene, can be expected to exist in these structures\upcite{carozo-NL-2011-raman}.

Raman spectroscopy is used to probe graphene samples, as one of the most useful and versatile tools\upcite{ferrari-PRL-2006-raman,Ferrari-nn-2013}. The measurement of the SLG, 2LG and FLG Raman spectra is a good way to understand the physical process, such as phonons, electron-phonon, magnetron-phonon and electron-electron interactions\upcite{Ferrari-nn-2013}. In bulk graphite, besides the G modes at 1582cm$^{-1}$, there are another two degenerate E$_{2g}$ modes at 42 cm$^{-1}$.\upcite{nemanich-1977-SSC-infrared,mani-pss-1974-lattice} Because the G band is an in-plane optical phonon and the interlayer coupling in graphite is very weak, therefore, there is no obvious difference between the G frequency in intrinsic (un-doped) mono-layer graphene, FLGs and graphite\upcite{Ferrari-nn-2013}. The E$^2_g$ mode at 42 cm$^{-1}$ in graphite corresponds to in-plane shear displacements between adjacent layers\upcite{Nemanich-1975-graphite,tan-2012-NM-shear}. Because this shear mode provides a direct measurement of the interlayer coupling, it is named as the C mode.\upcite{tan-2012-NM-shear} Because adjacent layers are rigidly displaced with respect to each other against the weak inter-layer restoring force, the C mode is not usually observed in typical Raman spectra for graphite samples due to its weak intensity and low frequency\upcite{hanfland-1989-prb-graphite,tan-2012-NM-shear}. Recently, using three Brag Grate notch filters (BNF) in combination with a single monochromator with high throughput, the C modes of BLG and FLGs were measured\upcite{tan-2012-NM-shear}. The experimental results agreed very well with a linear-chain model in theory\upcite{tan-2012-NM-shear}. Soon later, similar shear modes were found in other 2D materials, such as MoS$_{2}$\upcite{zhang-PhysRevB-2013} and WSe$_{2}$\upcite{zhao-2013-NL-interlayer}. Based on the symmetric analysis, for FLG with N layers, there are N-1 degenerate pairs of in-plane shear modes between neighbouring layers\upcite{tan-2012-NM-shear}. Up to now, only the C mode with the highest frequency was observed in FLGs. Because the other Raman-active modes has a much weaker intensity than the C peak as a result of a much smaller electron-phonon coupling, which had been confirmed by ab initio calculations performed using density functional theory\upcite{tan-2012-NM-shear}. It would be challenging to detect those shear modes.

In this paper, in contrast to that just one shear mode was observed in 2-4LGs, we observed all the shear modes of 2-4LGs in the Raman spectra of 2-4 layer scrolls (LSs), which are rolled up by the corresponding FLGs formed at the sample edges. The shear modes of 2-4LGs were measured for comparison. The observed frequency of all the shear modes agrees well with theoretical prediction of corresponding 3-4LGs. The results show a clear evidence that the layer coupling in FLGs will strongly modify the corresponding phonon spectrum in the lower frequency region, which has not been experimentally demonstrated clearly until now for such a layered system.

\section{Results and Discussion}
\begin{figure}
\centerline{\includegraphics[width=120mm,clip]{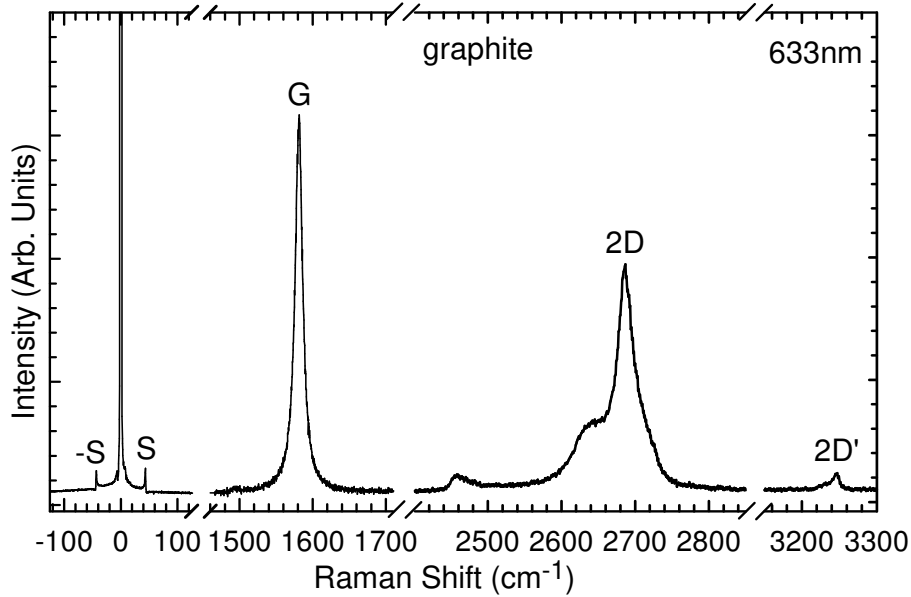}}
\caption{First- and second-order Raman spectra of bulk graphite excited by
633nm excitation. }
\label{fig:1}
\end{figure}

We first discuss the first- and second-order Raman modes of bulk graphite,
as shown in Figure 1. All the Raman modes in Figure 1 are with the same
scales of intensity and wavenumber. The asymmetrical peak at about
2700cm$^{-1}$ is the second order of the D peak, which is due to the
breathing modes of six-atom rings.\upcite{Ferrari-nn-2013} The peak at about 3250cm$^{-1}$ is the
second order of the D$'$ peak, which is at about 1620 cm$^{-1}$ in
defected graphite measured at 514 nm. The D and D$'$ bands require a
defect for their activation by intervalley and intravalley double
resonance Raman processes,\upcite{Thomsen-PRL-2000,Saito-PRL-2002,Tan-prb-2002} respectively. Therefore, they are absent in
pure graphite. In the lower
frequency region, a pair of Stokes and anti-Stokes modes are found at 43.5
cm$^{-1}$, which is close to the theoretical frequency of 42 cm$^{-1}$ for
the shear mode.\upcite{tan-2012-NM-shear} Here, the
simultaneous measurement of Stokes and anti-Stokes signal enables a
precise determination of the frequency of the shear mode. The FWHM of the
G peak in graphite is as large as 12.0 cm$^{-1}$, which is dominated by
its electron-phonon coupling strength because of its nature of
semi-metal.\upcite{lazzeri-PRB-2006-phonon} However, that of the shear mode is very
narrow, only 1.6 cm$^{-1}$. If excluding the instrument broadening of 0.6
$^{-1}$, which can be estimated from the broadening of laser excitation in
the Raman spectra, the intrinsic FWHM of the shear mode is as narrow as
1.0 cm$^{-1}$. As shown in Figure 1, the peak intensity of the shear mode,
I(C), is about 7\% of I(G), while the area intensity A(C) is about 0.6\%
of A(G) due to the narrow width of the shear mode. The weak intensity,
narrow width and low frequency of the shear mode make it difficult to be
observed in the single stage spectrograph with a normal notch (or edge)
filter and also in the triple monochromator system.\upcite{tan-2012-NM-shear}

\begin{figure}
\centerline{\includegraphics[width=120mm,clip]{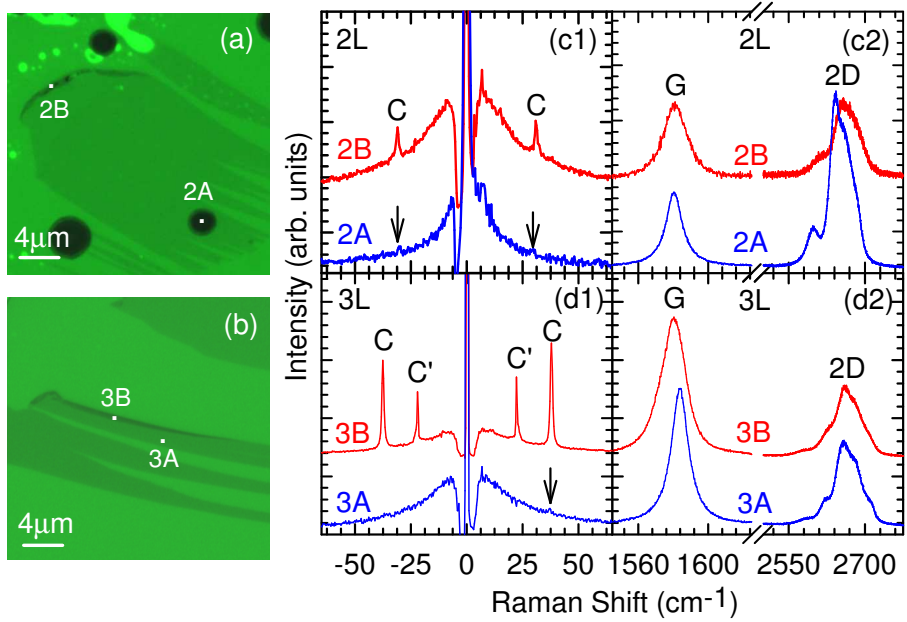}}
\caption{Optical images of 2LG (2A) and
its scrolls (2B) (a), and 3LG (3A) and its scrolls(3B) (b),
obtained using a 532 nm bandpass filter. Raman spectra of
2-3LGs (2A, 3A) and their scrolls (2B,3B) in the low frequency region (c1,d1) and G- and
2D-band region (c2,d2).}
\label{fig:2}
\end{figure}

Figure 2 shows the optical images of 2-3LGs and the corresponding 2-3LSs
and the related Raman spectra. Under the excitation of 633nm, the layer
number of suspended/supported 2-4LGs can be easily identified from the
line shape of the 2D peak. Indeed, the suspended 2LG and supported 3LG
located at position 2A and position 3A in Figures 2(c2) and 2(d2),
respectively, exhibit typical 2D peaks of 2LG and 3LG. In the low
frequency region, one Raman mode was observed for each 2LG and 3LG,
respectively, at 31 and 38 cm$^{-1}$. These modes could be assigned to the
shear mode in analogy to that in graphite according to their frequency
regions.\upcite{cardona-2007-Springer-light} Their frequency is prominently lower than
that (43.5 cm$^{-1}$) in graphite, as shown in Figure 1. Although its
intensity is very weak, it can be identified from the strong background as
indicated by arrows in Figure 2. FLGs could be incidentally rolled up to
form scroll structures at the their sample edge during the sample
preparation process. As examples, the wire-like scroll structures near the
position 2B and position 3B in Figures 2(a) and 2(b), respectively, show
the corresponding 2-3LSs. Indeed, the Raman spectra of 2-3LSs at position
2B and position 3B show similar 2D profiles of the 2-3LGs with a
blue-shift of several wavenumber in frequency, which is an analogy to the
case in the incommensurate 2LG. This can be interpreted as the reduction
of Fermi velocity at Dirac point.\upcite{ni-PRB-2008-reduction,kim-prl-2012-raman} The G mode of 2-3LSs still
exhibits a lorenzian line shape while its width is slightly broadened in
comparison to the G mode of 2-3LGs, however, the shear mode in 2-3LGs are
strongly enhanced in the 2-3LSs. In principle, because 2-3LSs are rolled
up from 2-3LGs, more graphenes layers in scrolls could significantly
increase the corresponding shear mode. We will discuss it in detail later.
Interestedly, an additional mode was observed at 22 cm$^{-1}$ in the 3LS.

\begin{figure}
\centerline{\includegraphics[width=120mm,clip]{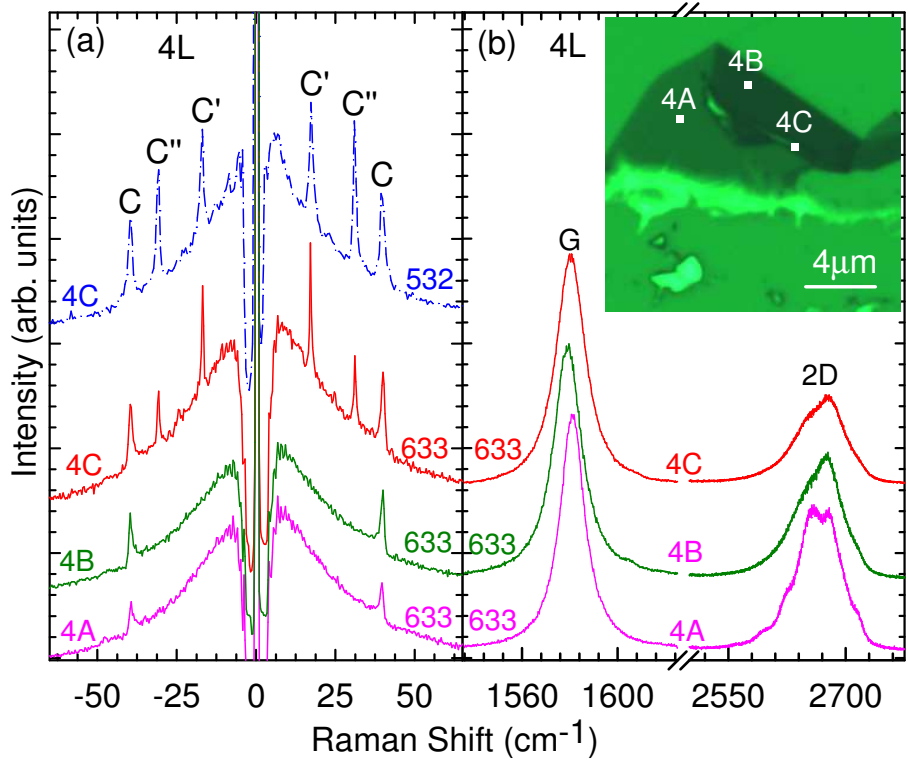}}
\caption{(Color online) Raman spectra of
4LG (4A), t4LG (4B) and 4LS (4C) in the low frequency region by 633nm and 532nm excitations (a) and G- and
2D-band region by 633nm excitation (b). The inset gives the optical image of 4LG and the corresponding scrolls and the t4LG obtained using a 532 nm bandpass filter. }
\label{fig:3}
\end{figure}

Figure 3 shows the optical image and Raman spectra of a 4LG and the
corresponding 4LS and the twisted 4LG (t4LG) folded by 4LGs. When the
633nm laser was used as an excitation, the Raman spectra at position 4A
shows a typical spectral feature of 4LG. A shear mode was observed at 40
cm$^{-1}$. The multiple structures of the 2D peak of 4LG was smeared out
in the t4LG at position 4B, while the intensity of the C mode at 40
cm$^{-1}$ is twice as much as that in 4LG at 4A. Similar with 3LS, the 2D mode in 4LS at
position 4C moves to higher frequency in comparison to that of 4LG in 4A
with a similar profile to 4LS in 4B. However, in the low frequency region,
two additional mode were observed at 17 and 31 cm$^{-1}$ in the 4LS. To
further confirm these Raman features, a 532nm laser was utilized for the
Raman spectra of the 4LS at 4C. Again, three modes were observed with
identical frequencies to those excited by the 633nm excitation,
respectively.

To reveal the nature of other modes observed in the low frequency region
of 3-4LSs, the optical modes of FLGs were calculated based on a force constant model\upcite{wang-2009-JRS-vibrational} and a linear chain model.\upcite{tan-2012-NM-shear} According to the symmetrical analysis of an $n$ ($n$$\geq$2) layer
graphene,\upcite{malard-PRB-2009-group,wang-2009-JRS-vibrational} there are $n$ degenerate pairs of in-pane
stretching modes and $n$-1 degenerate pairs of in-plane shear modes
between neighboring layers. The former corresponds to the feature in the
G-band region and the latter appears in the low frequency region. For the
2-4LGs, in principle, there are 1, 2 and 3 degenerate pairs of the shear
mode, which is consistent with the experimental observation. The
force-constant model\upcite{wang-2009-JRS-vibrational} is adopted to calculate the frequency of
the shear mode and the G band, in which 15 parameters are used to describe
fifth-nearest neighbor force constants of intra-layer and 8 parameters for
fourth-nearest-neighbor force constants of interlayer. The theoretical
values for the two Raman-active degenerate E$_{2g}$ modes of graphite are,
respectively, at 1581.7 and 42.0 cm$^{-1}$. The former is the in-plane
stretching mode, i.e., the G mode. The latter is the in-plane shear mode
between adjacent graphene layers in graphite. Their theoretical
frequencies are consistent with the experimental ones. Figure 4 shows the
normal mode atomic displacements and the corresponding symmetries of the
shear modes for 2-4LGs and bulk graphite at $\Gamma$ point. The
theoretical frequency of all the shear modes are summarized in Table I.
The frequency of the shear mode observed in 2-4LGs agrees well with that
of the shear mode between adjacent two graphene layers, similar to the
atomic displacements of the shear mode in graphite observed at 43.5
cm$^{-1}$. Both theoretical and experimental results show that such C mode significantly
reduces in frequency with decreasing the layer number. The theoretical
values of the other shear modes, i.e., E$'$ mode in 3LGs, and E$_u$ and
E$^1_g$ modes in 4LGs, match well with those of additional modes observed
in 3-4LSs in Figures 2(d1) and 3(a). All the graphene layers in 2-4L
scrolls is hard to keep the commensurate bernal stacking structure, which is hard
to be kept in the incidental rolling process to form the scroll
structures. Therefore, it is reasonable that all the theoretical shear
modes in 2-4LGs are observed in 2-4LSs with almost the same frequency. Their small discrepancy could be attributed to theoretical error and additional strain existed in 2-4LSs.
The additional Raman active shear modes in 3-4LGs were denoted as the C$'$ mode, and the additional
infrared-active mode in 4LGs as the C$''$ mode in this work. All the
observed shear modes in 2-4LSs are labeled with corresponding denotations
in Figures 2(d1) and 3(a). The observed C$'$ modes of 3-4LSs are,
respectively, attributed to the shear mode of two outer layers with
respect to center layer(s) of 3-4LGs, and the C$''$ mode of the 4LS is
attributed to the shear mode between top and bottom two layers of the 4LG. Table I also includes the theoretical results of the frequencies of the shear modes based on the linear chain model\upcite{tan-2012-NM-shear}, which also agree well with the experimental results.

\begin{figure}[tb]
\centerline{\includegraphics[width=120mm,clip]{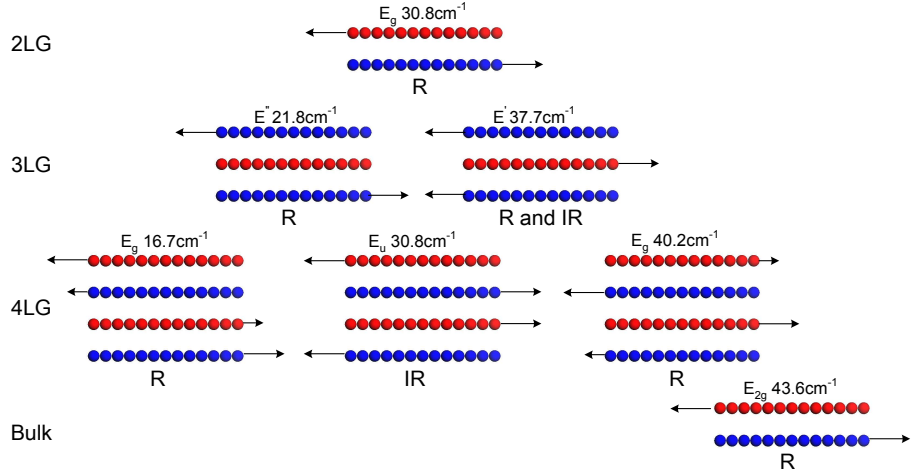}}
\caption{(Color online) Normal mode atomic displacements of the shear modes for 2-4L graphenes and bulk graphite. The symbol under each mode
is its irreducible representation. R or IR indicate if the mode is Raman, or infrared active, or both. The theoretical results based on the force constant model are also indicated.}
\label{fig:4}
\end{figure}

Many graphene scrolls have a wire morphology in the local area at the sample edge. Raman spectra with parallel (XY) and perpendicular (YY)
polarizations of the configuration of incident and scattered light were collected upon the 3LS in Figure 2(b). The optical image and polarization configuration were shown in Figure 5(a). We recorded the backscattering Raman signal with polarization along the Y direction. A half-wave plate was used to change the polarization direction of the incident excitation $\vec{e}_i$ along either parallel (Y) or perpendicular (X) to the Raman signal. Raman images of (XY) and (YY) are shown in Figures 5(b1) and 5(b2). The peak intensity of the C and C$'$ modes is, respectively, displayed in the images as green and red colors. Therefore, both the intensity and intensity ratio of the C and C$'$ modes are demonstrated in images as the brightness with different colors. As an example, Figures 5(c1) and 5(c2) show Raman spectra at different spots in Figure 5(b2). The images show that the peak intensity and polarization of the C and C$'$ modes, and intensity ratio between the C and C$'$ modes are dependent on the measured position. The C mode is too weak to be observed for supported 3L graphenes and for the top and bottom parts of 3LSs. Therefore, the strong C and C$'$ modes in scrolls could not be related to more graphene layers in 3LSs. It should be dependent on the local structure of graphene scrolls. Similar results were observed in other graphene scrolls.

\begin{table}
\caption{Theoretical (Theo.) values (in cm$^{-1}$) of the shear modes in 2-4LGs based on the force constant model (FCM) and linear chain model (LCM), and the corresponding experimental (Exp.) ones in 2-4LGs and the corresponding scrolls (2-4LSs). The notation of each observed shear mode of 2-4LSs in Figures 2 and 3 are indicated within parentheses after the experimental values.}
\begin{center}
\begin{tabular}{c|c|c|c|c|c|c|c}
\hline
&2L&\multicolumn{2}{c|}{3L}&\multicolumn{3}{c|}{4L}\\\hline
Mode&E$_g$(R)&E$''$(R,IR)&E$'$(R)&E$^2_g$(R)&E$_u$(IR)&E$^1_g$(R)\\\hline
Theo.(FCM)&32.0&37.8&23.4&39.7&32.0&18.2\\\hline
Theo.(LCM)&30.8&37.7&21.8&40.2&30.8&16.7\\\hline
Exp. in 2-4LGs&30.0&37.2&-&39.2&-&-\\\hline
Exp. in 2-4LSs&31.2 (C) &37.6 (C)&21.9(C$'$)&39.5(C)&30.7(C$''$)&16.7(C$'$)\\\hline
\end{tabular}
\end{center}
\end{table}

In principle, the C$''$ mode observed in 4LSs in Figure 3 (i.e., the E$_u$ mode of 4L graphenes in Figure 4) is infrared active and Raman inactive. Indeed, it is absent in supported 4LG. Both the C$'$ modes of 3-4LSs in Figures 2(d1) and 3(a) are Raman active, however, they are also absent in supported 3-4LGs. The absence of the C$'$ mode in 3-4LG may result from their weak Raman cross section. In comparison to the C mode, the C$'$ mode should be very weak or unobservable in 3-4LSs if without any resonance in 3-4LSs. However, the C$'$ mode can be stronger than the C mode in 3-4LSs as indicted in Figures 3(a) and 5(c1). The modification of crystalline point-group symmetry in scrolls can relax wave-vector selection rule by loss of translational symmetry and make infrared C$''$ mode Raman-active, similar to the observation of Raman-inactive B$_{2g}$ mode at 867 cm$^{-1}$ at the edge plane of graphite.\upcite{kawashima-RRB-1999-observation} However, its strong
intensity in the 4LS is still an unexpected result. Furthermore, the C and C$'$ mode intensity in the 3LS is independent on the its optical contrast, i.e., the number of graphene layers of the 3LS. All the aspects point to that all the shear modes in 2-4L scrolls were resonantly enhanced in some local positions so that the Raman-inactive or absent Raman modes in supported 3-4LGs become visible in 3-4L scrolls.

The resonant enhancement of the C modes in 3-4L scrolls reveals that their electronic structure near Dirac K point are quite different from those of the corresponding 3-4L graphenes. Recently, two symmetric low-energy Van Hove singularities in the density of states in twisted graphene bilayers (t2LG) were observed by scanning tunneling spectroscopy\upcite{li-2009-NPhysi-observation}. Such position of these singularities can be tuned by controlling the relative angle between layers. A rotation between stacked graphene layers can generate Van Hove singularities, which can be brought arbitrarily close to the Fermi energy by varying the angle of rotation. The Raman G band intensity is strongly enhanced due to the singularities in the joint density of states of t2LG, whose energy is exclusively a function of twist angle and whose optical transition strength is governed by interlayer interactions\upcite{havener-NL-2012-angle,kim-prl-2012-raman}. In the 2-4LSs, there would be many different twist-angle local staking configurations between each 2-4LG, and some of their singularity energies may resonant with the energy of phonons corresponding with shear modes. This enhancing of all shear modes in 2-4LGs is different with that of G band, which is from the resonance between energy of excitation and singularity energy. This is the reason why the frequencies of shear modes are identical by 633nm and 532nm excitations, shown in figure 3(a), and the enhancing of G band was not observed in any samples.

\begin{figure}[tb]
\centerline{\includegraphics[width=120mm,clip]{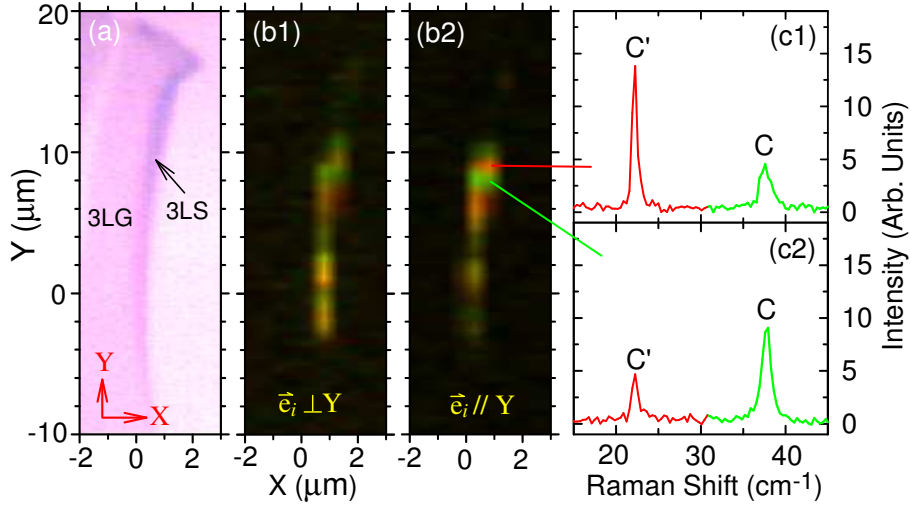}}
\caption{Optical image (a) of the 3L graphene and
its scrolls and the corresponding pseudo-color Raman image (b1,b2) by 633nm
when the laser polarization $\vec{e}_i$ is, respectively, parallel
and perpendicular to the Y axis, where the red and green colors reflect
the Raman intensity of the C$'$ (22cm$^{-1}$) and C (38cm$^{-1}$) modes,
respectively. Two typical Raman spectra in the image of (b2) are shown
in (c1,c2).}
\label{fig:5}
\end{figure}

Moreover, the first principle calculation shows that a carbon scrolls rolled up by a single graphene sheet can be metallic, semi-metallic or semiconducting, which is greatly dependent on their chirality\upcite{chen-JPCC-2007-structural}. The semiconducting scrolls have an energy gap dependent on their sizes and chirality, which can be as large as several 100 meV. E. J. Mele found that structures in t2LG that are even under sublattice exchange are generically gapped in theory\upcite{mele-2010-PRB-commensuration}. The random rolling of 2-4LGs to form 2-4LSs produces different local staking configurations between each 2-4LG, so both nanostructures, which result in different local electronic structures, mentioned in Chen's\upcite{chen-JPCC-2007-structural} and Mele's\upcite{mele-2010-PRB-commensuration} work likely exist in 2-4LSs. According to the previous theoretical calculations, the band gap opening is expected in the 2-4LSs dependent on the local staking configurations of 2-4LGs inside the 2-4LSs. The shear modes, such as the C and C$'$ modes in 3-4LSs and the C$''$ mode in 4LSs could be resonantly enhanced by such small band gaps. Similar with the singularities energy, the band gaps dependent on the local staking configurations, which make the intensity of shear modes is different on different position, as shown in Fig5.

\section{CONCLUSIONS}
In conclusion, the one, two and three low frequency modes below 50.0 cm$^{-1}$ were observed in the Raman spectra of graphene scrolls of 2-4 layers, which are attributed to the Raman- and infrared-active in-plane shear mode between neighbor-layers of graphene 2-4 layers according to the theoretical results based on the force-constant and linear chain models. All the shear modes are with a narrow line width less than about 1.4 cm$^{-1}$. The incommensurate Bernal stacking structures in 2-4LSs are the reason why other shear modes, which can't be observed in well-stacking 2-4LGs, were measured. In addition, the intensity of shear modes are dependent on measured position. The local staking configurations, which decide the local electronic structures, distribute randomly in 2-4LSs, so the intensity enhancement of shear modes form particular local electronic structures is spatially different. This observation of other shear modes in 3-4LSs reveals the inter-layer interaction in 2D materials in detail and comprehensively, and more shear modes are excepted to be measured in more layer (layer number $>$ 4) graphene scrolls. This all-scale shear modes observation can be expected to happen in other 2D materials with similar scroll structures.

\section{Experimental Section}
In this work, graphene samples were obtained by micro-mechanical cleavage
of natural graphite on the surface of a Si wafer chip with 293 nm or 92nm
thick SiO$_2$ on the top.\upcite{novoselov-NAS-2005-two} 2-4L graphenes with a bernal
stacking were identified by Raman spectra from their spectral profiles of
the 2D mode. Raman measurements were performed in a backscattering
geometry at room temperature using a Jobin-Yvon HR800 Raman system. The
excitation wavelengths are 633nm of a He-Ne laser and 532nm of a diode
pumped solid-state laser. The laser excitations were cleaned from their
plasma lines with BragGrate Bandpass filters. Measurements of Raman modes
with frequencies close to 5 cm$^{-1}$ were enabled by three BragGrate
notch filters at 633 nm or 532nm with optical density 3 and with the full
width at half maximum (FWHM) of 5-10 cm$^{-1}$. All the Raman measurements
were done using a 100X objective lens(NA=0.90) and a 1800 lines/mm grating
to get a spectral resolution 0.35 cm$^{-1}$ per CCD pixel. The typical
laser power was used 0.5mW in order to avoid sample heating.\upcite{tan-1999-APL-intrinsic}

\section{Acknowledgment}
This work was supported by the National Basic Research Program of China (973 Program)
Grant No. 2009CB929301, and National Natural Science Foundation of China under grant Nos.~10934007 and 11225421.

\bibliographystyle{small}
\bibliography{myreference}


\end{document}